# Interplay between magnetism and superconductivity and appearance of a second superconducting transition in α-FeSe at high pressure.

V.A. Sidorov[1], A.V. Tsvyashchenko[1], R.A. Sadykov[1,2]

[1]Institute for High Pressure Physics, Russian Academy of Sciences, 142190 Troitsk, Russia
[2]Institute for Nuclear Research, Russian Academy of Sciences, 142190 Troitsk, Russia

We synthesized tetragonal α-FeSe by melting a powder mixture of iron and selenium at high pressure. Subsequent annealing at normal pressure results in removing traces of hexagonal β-FeSe, formation of a rather sharp transition to superconducting state at $T_c \sim 7$ K, and the appearance of a magnetic transition near $T_M = 120$ K. Resistivity and ac-susceptibility were measured on the annealed sample at hydrostatic pressure up to 4.5 GPa. A magnetic transition visible in ac-susceptibility shifts down under pressure and the resistive anomaly typical for a spin density wave (SDW) antiferromagnetic transition develops near the susceptibility anomaly. $T_c$ determined by the appearance of a diamagnetic response in susceptibility, increases linearly under pressure at a rate $dT_c/dP = 3.5$ K/GPa. Below 1.5 GPa, the resistive superconducting transition is sharp; the width of transition does not change with pressure; and, $T_c$ determined by a peak in $d\rho/dT$ increases at a rate $\sim 3.5$ K/GPa. At higher pressure, a giant broadening of the resistive transition develops. This effect cannot be explained by possible pressure gradients in the sample and is inherent to α-FeSe. The dependences $d\rho(T)/dT$ show a signature for a second peak above 3 GPa which is indicative of the appearance of another superconducting state in α-FeSe at high pressure. We argue that this second superconducting phase coexists with SDW antiferromagnetism in a partial volume fraction and originates from pairing of charge carriers from other sheets of the Fermi surface.

**Introduction**

Recent discovery of high $T_c$ superconductivity in layered iron arsenides [1-4] extended the family of unconventional superconductors. Previous work on high $T_c$ cuprates and heavy fermion superconductors prepared a basis for rapid progress in research on these new materials. The proximity to a magnetic instability, complex gap function, and coexistence of magnetism and superconductivity provide a key framework for future understanding the pairing mechanism of iron-based superconductors. The common structural feature of all these materials is layers composed of edge-sharing $FeAs_4$-tetrahedra, separated by rare-earth-oxygen or alkaline-earth layers. The tetragonal α-phase of iron selenide has a structure composed of a stack of edge-sharing $FeSe_4$-tetrahedra layer by layer, without any additional separating elements and may be regarded as an end member of a series of iron-based superconductors. Therefore, the discovery of superconductivity in α-FeSe with $T_c = 7$ K [5] attracted considerable attention. Specific heat [5] and NMR [6] measurements indicate the unconventional nature of superconductivity in α-FeSe with lines of vanishing gap on the Fermi surface. In spite of the relatively simple structure of this binary compound, the preparation of a single phase sample is a challenge. Standard solid state reaction usually results in two or more phases in the sample, the impurity hexagonal β-phase being dominant [5-9]. In the first publications on superconductivity in iron selenide, the samples were off-stoichiometric, with a small deficiency in selenium, namely $FeSe_{0.88}$ [5,7]. Further studies revealed that the highest superconducting transition temperature and a nonmagnetic ground state are inherent to nearly stoichiometric, single phase $Fe_{1.01}Se$ [10]. There is some controversy in the designation of phases of FeSe. The authors of Ref.10 refer to the tetragonal phase as β-$Fe_{1.01}Se$ and the hexagonal phase as α-FeSe, in accordance with designation made in early publications; while most others (see, for example Ref. 6-8) call the tetragonal phase α-FeSe. We also refer to the tetragonal FeSe phase as α-FeSe. While stoichiometric FeSe is non magnetic [10], Se-deficient superconducting samples exhibit magnetic transitions of still unknown origin [7,8]. Band structure calculations of Lee et al. [11] show that Se-deficiency drives FeSe close to magnetism and short-range magnetic correlations in this material are rather strong. Recently, Imai et al. [12] performed NMR measurements on $Fe_{1.01}Se$ under 2 GPa pressure and found an enhancement of antiferromagnetic spin fluctuations in parallel with a rise of $T_c$. This may be indicative of the important role of magnetism to superconductivity in FeSe. Band structure calculations of Subedi at al. [13] reveal that FeSe has cylindrical Fermi surfaces, both for electrons and holes (similar to iron arsenides), which satisfy nesting conditions that result in a spin density wave (SDW) instability and appearance of an antiferromagnetic ground state. So, both experiments and calculations show that there is a subtle interplay between superconductivity and magnetism in tetragonal FeSe, which makes it very close to iron arsenide superconductors. High pressure experiments up to 1-1.5 GPa [7,9] demonstrated a pronounced increase of $T_c$ with applied pressure. Mizuguchi et al. claimed a huge $dT_c/dP$ = 9.1 K/GPa and $T_c$ onset of 27 K at 1.5 GPa [9]. High pressure studies on iron arsenide superconductors (for a review see paper of C.W. Chu and B. Lorenz [14]) reveal complex behavior, where $dT_c/dP$ may be positive and negative, depending on pressure and doping. These publications motivated us to study FeSe at pressures higher than 1.5 GPa.

**Experiment**

To prepare samples we mixed powders of iron and selenium in a desired stoichiometry put them in a sleeve made of a NaCl single crystal, and pressed the powders in a "toroid" high pressure apparatus to a few GPa. The powder mixture was heated by a directly passing current through it until melting was fixed via the observation of an anomaly on the current-voltage dependence. The heating power was then removed. The overall synthesis procedure took a few minutes. This method has been described earlier [15] and was applied for the synthesis of many rare-earth compounds under pressure. High pressure synthesis of iron selenide was most successful near 5 GPa, where a maximum on the melting curve of selenium takes place. In the process of high pressure synthesis in the closed cell, the initial composition of the sample can not change as the components have no chance to evaporate or leave the sample in some other way. Samples of two nominal compositions FeSe and FeSe$_{0.88}$ were prepared. They were checked for phase purity by x-ray powder diffraction (IPDS STOE diffractometer equipped with an image plate detector). We used Mo K$_\alpha$ radiation because elemental iron gives a much better diffraction on Mo radiation than on Cu. Electrical resistivity and ac-susceptibility measurements were performed on pieces of bulk polycrystalline samples using a lock-in detection technique (SR830 lock-in amplifier and SR554 transformer-preamplifier). High pressure experiments were performed with a small clamped "toroid" device. A Teflon capsule 2 mm in diameter and 2 mm in height was filled with a glycerol-water mixture (3:2 in volume) which served as a hydrostatic pressure medium around the samples. It solidifies above 5.3 GPa at room temperature. Other details of the high pressure technique can be found elsewhere [16]. Four Pt wires 25 μm in diameter were spot-welded to the sample for resistivity measurements. A sample for ac-susceptibility measurements at high pressure was placed in a small coil system (0.7 mm in diameter and 0.8 mm in length, having 25 turns in both primary and secondary coils) together with a small chip of Pb which served as a pressure sensor. Pressure dependence of the superconducting transition temperature of Pb was used to calculate pressure according to Eiling and Schilling [17]. Susceptibility measurements at ambient pressure were done with a SR830 lock-in amplifier in a large compensated coil system, but pressure measurements of ac-susceptibility in a small coil were possible only with a SR554 preamplifier. The measurement frequency was 157 Hz and current through the primary coil was 5 mA in this case. Temperature was measured with a calibrated silicon diode with a resolution 10 mK.

**Results and discussion**

X-ray diffraction patterns of as-prepared samples are shown in the two upper panels of Fig.1. Most peaks in the patterns belong to tetragonal α-FeSe. The remaining peaks may be indexed as the strongest peaks of hexagonal β-FeSe (labeled by asterisks (*)) and elemental α-Fe (labeled by an open circle). The hexagonal phase presents as an impurity in both samples in approximately equal proportions. On the other hand, elemental iron was detected only for a sample with the initial composition $FeSe_{0.88}$. This means that high pressure is favorable for the synthesis of a more stoichiometric sample and that ~12% excess iron put in the starting composition does not conform to the lattice of FeSe and forms iron clusters. Probably the formation of Se vacancies at high pressure is energetically less profitable, and the resulting iron selenide crystals formed under pressure are nearly stoichiometric. Lattice constants of the tetragonal unit cell of an annealed sample "$FeSe_{0.88}$", determined from fitting of peak positions, are a = 3.772 Å, c = 5.525 Å, c/a = 1.4647. These values are comparable with data published by others, but most closely conform to the values published for $Fe_{1.01}Se$ (a = 3.7734 Å, c = 5.5258 Å, c/a = 1.4644) [10].

The temperature dependences of electrical resistivity of both high pressure synthesized samples are shown in Fig.2. The room temperature value of resistivity 25-30 mΩ-cm is an order of magnitude higher than that reported for samples synthesized by solid state reaction at ambient pressure (2-3 mΩ-cm). It may indicate a high degree of imperfection and defects in a crystal lattice. This may be due to high cooling rates in the high pressure synthesis process. Temperature dependences of both samples become practically indistinguishable below 20 K and they exhibit a pronounce drop from ~ 22 mΩ-cm to 2 mΩ-cm between 8 and 1.7 K, which signifies the formation of superconducting states below ~8 K. However, ac-susceptibility measurements did not show a diamagnetic response or any other magnetic anomalies in the entire range of temperatures 1.7-300 K. This means that superconductivity below 8 K in these as-prepared samples is not bulk, but rather filamentary.

We annealed our samples at normal pressure. The samples were put in quartz ampoules and then a procedure of evacuation and flushing with pure argon gas was repeated a few times. Finally, the ampoules were filled with argon and placed in the oven at 400°C for 34 h. The x-ray diffraction patterns of annealed samples show the disappearance of traces of hexagonal β-phase and better resolution of all peaks, indicating a more equilibrium and less strained crystal lattice of the remaining α-phase (bottom panel of Fig.1). Ac-susceptibility measurements on annealed samples exhibit signatures of superconductivity below 7 K and an additional small feature near 120 K. Fig.3 displays the dispersive and dissipative parts of ac-susceptibility of the annealed "FeSe$_{0.88}$" sample (its composition is really close to FeSe). Very similar results were obtained for an annealed "FeSe" sample. In the region of the superconducting transition, the dissipative part $\chi''$ of ac-susceptibility exhibits a broad peak around the temperature where the dispersive part $\chi'$ has a maximum temperature derivative. These features are typical for many other superconductors. The anomaly in ac-susceptibility at $T_M$ = 120 K may indicate the formation of a magnetic order below this temperature. The amplitude of the anomaly is very small, thus precluding the formation of ferromagnetic order. Magnetic anomalies in this temperature range were observed by others in off-stoichiometric samples [7,8]. Moreover, specific heat anomalies indicative of bulk magnetic transitions, were also reported for FeSe$_{0.88}$ [7]. However, the origin of magnetism in FeSe is still unknown. The room temperature resistivity of the "FeSe$_{0.88}$" sample decreases 10 times after annealing and the resistive transition to a superconducting state becomes rather sharp. Unfortunately, the "FeSe" sample was probably slightly oxidized in the process of annealing, and a sharp superconducting transition at 7 K in this sample was superimposed on a big non-superconducting background resistance. Probably oxidation occured only at the inter-crystallite layers. Therefore, we observed very similar behaviors of the ac-susceptibility in both annealed samples and different behaviors of their resistivity.

Pressure measurements were performed on two pieces taken from the annealed "FeSe$_{0.88}$" sample. Fig.4 displays the temperature dependences of resistivity ρ(T) of FeSe at different pressures. Two main features of these dependences are of interest. The superconducting transition temperature shifts up in temperature and the transition becomes broad at high pressure. The $T_c$ onset temperature reaches 36.5 K at 4.55 GPa, formally giving an average $dT_c/dP$ = 6.1 K/GPa in this pressure range. Another interesting feature is a gradual appearance of an anomaly in ρ(T) at high pressure, which is characteristic for magnetic transitions of the SDW type. With increase of pressure, the anomaly is downshifted in temperature as indicated by a blue dashed arrow in Fig.4. Extrapolation to ambient pressure implies that the anomaly would be located around 120 K. At this characteristic temperature $T_M$, an ac-susceptibility anomaly is clearly seen (upper panel of Fig.3) but the $\rho(T)$ dependence only has an inflection point. There is a strong analogy with the evolution of $\rho(T)$ dependences of FeSe under pressure plotted in Fig.4 and those observed recently by Kotegawa et al. [18] for single crystals of $SrFe_2As_2$, which is a SDW antiferromagnet below $T_M$ = 198 K but becomes a superconductor under pressure above ~3.4 GPa. High pressure decreases $T_M$ in this material and magnetism is completely suppressed above 3.7 GPa. Very similar evolution of the shape of anomalies in $\rho(T)$ was observed by Fei Han et al. [19] for $SrFe_2As_2$ doped with Rh to the Fe sites. They observed the suppression of SDW antiferromagnetism by Rh doping, and appearance of superconductivity with $T_c$ up to 22 K. On increasing Rh doping, the resistance anomaly at $T_M$ transforms from "Fisher-Langer" type [20] (sharp drop of $\rho(T)$ at the transition, $d\rho(T)/dT$ has a peak at $T_M$) to "Suezaki-Mori" type [21] (increase of $\rho(T)$ at the transition due to a partial gapping of the Fermi surface). Thus, the magnetism in FeSe evolves under pressure in a way similar to that observed in another iron-based superconductor $SrFe_2As_2$ at high pressure and with Rh doping. Fig.5 shows a shift of the ac-susceptibility anomaly in FeSe, associated with a magnetic transition at $T_M$. Again, the application of high pressure decreases $T_M$.

Now we discuss the superconductivity of FeSe at high pressure in more detail. Fig.6 shows the response of a secondary coil near the superconducting $T_c$, which is proportional to ac-susceptibility + a nearly constant background. The superconducting transition temperature $T_c$ determined by the appearance of a diamagnetic response in susceptibility increases linearly under pressure at a rate $dT_c/dP$ = 3.5 K/GPa. The diamagnetic response appears at a temperature where the sample resistance approaches zero (Fig.7). The shape of $\rho(T)$ dependences at the superconducting transition changes appreciably with the application of pressure as shown in Fig.7. At pressures up to 1.5 GPa, the resistive anomaly shifts rigidly at a rate ~3.5 K/GPa similar to that observed in susceptibility measurements. But a huge broadening occurs at higher pressure. Broadening manifests itself as a stepped-up increase of the $T_c$ onset temperature. This effect cannot be explained by possible pressure gradients in the sample and is inherent to α-FeSe. The width of the superconducting transition temperature of Pb in this experiment is less than 10 mK and thus estimated pressure gradients across the sample are less than 0.03 GPa. The width of the superconducting transition in FeSe exceeds 12 K at 4.55 GPa and is certainly far beyond that produced by a 0.03 GPa pressure gradient. We can also preclude the irreversible deterioration of sample quality produced by high pressure. Measurements on unloading at 0.6 GPa and at ambient pressure after complete unloading of the pressure cell showed that the superconducting transition became narrow and is observed at the same temperature as before the application of high pressure. The 120-K anomaly in susceptibility is again at the same temperature on the unloaded sample. Careful inspection of $\rho(T)$ data shown in Fig.7 allows one to conclude that there is not a single superconducting transition above 3 GPa, but that additional superconducting states appear above the $T_c$ onset temperature of the main superconducting transition, distorting the shape of the resistive anomaly at $T_c$. The main transition continues to increase at high pressure with the same rate, but $T_c$ of the additional superconducting states increases with a higher rate, thus producing an effective giant broadening of the transition. Numerical differentiation of $\rho(T)$ confirms this conclusion (Fig.8). Initially one observes a single peak in $d\rho(T)/dT$ which is shifted without broadening up to 1.5 GPa, but broadens appreciably at 2.46 GPa (upper panel of Fig.8). At pressures above 3 GPa, $d\rho(T)/dT$ is approximated by two Gaussian peaks (three lower panels of Fig.8) which are shifted to high pressure with different rates. The lower temperature peak 1 continues to increase with the same rate as the single peak below 1.5 GPa, roughly 3.5 K/GPa in parallel with $T_c$ determined by ac-susceptibility (Fig.9). The higher temperature peak 2 increases faster, resulting in a broadening of the transition via an accompanying increase of the $T_c$ onset temperature. Why does the resistance of FeSe not go to zero when the higher temperature superconducting state appears and what is the origin of this state? Unambiguous answer on this question requires further experiments at high pressure. At the moment we can argue based on resistivity and susceptibility measurements that the new superconducting states do not spread out to the whole volume of the sample but are spatially located in some places. It is also possible the appearance of conditions for pairing at high pressure on some portions of Fermi surface that were connected initially by an SDW Q-vector in k-space, but not on the whole sheet of the Fermi surface, and the gradual evolution of this region in k-space away from perfect nesting with pressure. An analogy to this effect among iron-based superconductors can be found in $AEFe_2As_2$ (AE = Ca, Sr, Ba) for which µ-SR measurements on superconducting samples [22,23] reveal the existence of a static magnetic order coexisting with superconductivity in a partial volume fraction. This phenomenon was observed on different samples for which superconductivity was accessed by pressure or chemical substitution-induced tuning of the magnetic ground state.

Multiple superconducting transition temperatures observed in FeSe imply multiple superconducting gaps. Multiple gaps are possible in the case where charge carriers from different parts of a complex Fermi surface participate in superconducting pairing. The most well-known example is $MgB_2$ in which two gaps 2.5 and 7 meV were found by point contact spectroscopy (PCS) [24,25]. Two superconducting gaps also were observed by PCS in iron oxypnictides $LaFeAsO_{1-x}F_x$ and $SmFeAsO_{1-x}F_x$ [26] and iron arsenides $Ba_{1-x}K_xFe_2As_2$ and $Ba(Fe_{1-x}Co_x)_2As_2$ [27]. A multiband model for superconductivity of iron-based superconductors was considered by Benfatto et al. [28]. Typically, both gaps vanish at a single temperature $T_c$ as in $MgB_2$, $SmFeAsO_{1-x}F_x$ and $Ba_{1-x}K_xFe_2As_2$. However, they also can vanish at different temperatures as in $LaFeAsO_{1-x}F_x$ [26]. Note that multiple superconducting gaps are observed in hole-doped iron oxypnictedes and arsenides ($LaFeAsO_{1-x}F_x$, $SmFeAsO_{1-x}F_x$, $Ba_{1-x}K_xFe_2As_2$) but the electron-doped $Ba(Fe_{1-x}Co_x)_2As_2$ exhibits a single gap. The features of band structure and Fermi surface topology of all iron-based superconductors (including FeSe) are favorable for SDW magnetism and interband pairing, which may be closely related.

The P-T diagram of magnetism and superconductivity in FeSe up to 5 GPa is shown in Fig.10. SDW magnetism is suppressed under pressure and may completely disappear above ~ 8 GPa. Superconductivity, observed at normal pressure, is enhanced at high pressure at a rate $dT_c/dP$ = 3.5 K/GPa. Above ~1.5 GPa, additional superconducting states appear that result in a faster increase of the $T_c$ onset temperature with pressure. However, this fast increase of the $T_c$ onset temperature saturates near 5 GPa at a level ~ 37 K.

Very recently Margadonna et al. [29] and Medvedev at al. [30] presented results of resistance measurements and synchrotron x-ray diffraction on nonmagnetic $Fe_{1.01}Se$ at pressures up to 14-38 GPa. They found a very strong interlayer compressibility of FeSe at high pressure that may cause a strong enhancement of $T_c$ in this material. Above ~9 GPa, the tetragonal α-FeSe transforms to a denser β-FeSe, which is nonmagnetic and semiconducting. Medvedev et al. [30] also performed Mössbauer spectroscopy studies and showed that their samples have no static magnetic order in the range of temperature 4.2-300K up to 38 GPa. Both authors found by resistance measurements that the $T_c$ onset temperature of superconductivity passes through a maximum at ~37 K near 6-8 GPa.

The later result is in agreement with our observations. In both papers, the resistance of the sample was measured using a solid pressure medium NaCl [29] or cubic-BN/epoxy [30]. Medvedev et al. indicate that pressure gradient across the sample was less than 0.05 GPa in their resistance measurements, and the resistance vs temperature curve at 5.1 GPa, shown in Fig.2c of their paper [30], looks similar to our data at 4.55 GPa. Probably the broadening observed by Medvedev et al. is mainly due not to pressure gradients, but interpretation is difficult in this case. However, a giant broadening of the resistive superconducting transition was observed also for iron oxypnictide $LaFeAsO_{1-x}F_x$ by Takahashi et al. [31]. Below 3 GPa they used a piston-cylinder cell with a liquid pressure medium. The insert in Fig2a of their paper [31] is very similar to data in our Fig.9. Takahashi et al. proposed that this giant broadening was due to sample and stress inhomogeneities, but this explanation seems artificial, taking in mind that the broadening of the transition at their hydrostatic pressure is larger than the shift of the zero-resistance temperature at 3 GPa (stresses should exceed 3 GPa, which is improbable). Note again that multiple gaps vanishing at different temperatures were observed in $LaFeAsO_{1-x}F_x$ by Gonnelli et al. [26]. It looks plausible that a giant broadening of the resistive superconducting transition observed under hydrostatic pressure in $LaFeAsO_{1-x}F_x$ and FeSe has a common origin – multiple superconducting gaps that intervene at different pressures and temperatures.

**Conclusions**

We constructed the P-T diagram up to 5 GPa of magnetism and superconductivity for pressure synthesized samples of tetragonal FeSe. SDW magnetism is suppressed at high pressure and the superconducting transition temperature increases at a rate $dT_c/dP = 3.5$ K/GPa. The signature of other higher temperature superconducting states was found that may coexist with SDW magnetism. These states may be due to charge carriers from parts of the Fermi surface that no longer satisfy nesting condition for SDW order. We propose that multiple gap structures may be the origin of a giant broadening of the resistive superconducting transition observed at hydrostatic pressure in other iron-based superconductors.

**Acknowledgements**

This work was supported by the Russian Foundation for Basic Research (grant 07-02-00280), Program of the Division of Physical Sciences of RAS "Strongly Correlated Electrons" and Program of the Presidium of RAS "Physics of Strongly Compressed Matter". We are grateful to S.M. Stishov for support of this work and discussion of the results and Joe D. Thompson for helpful discussions and reading the manuscript.

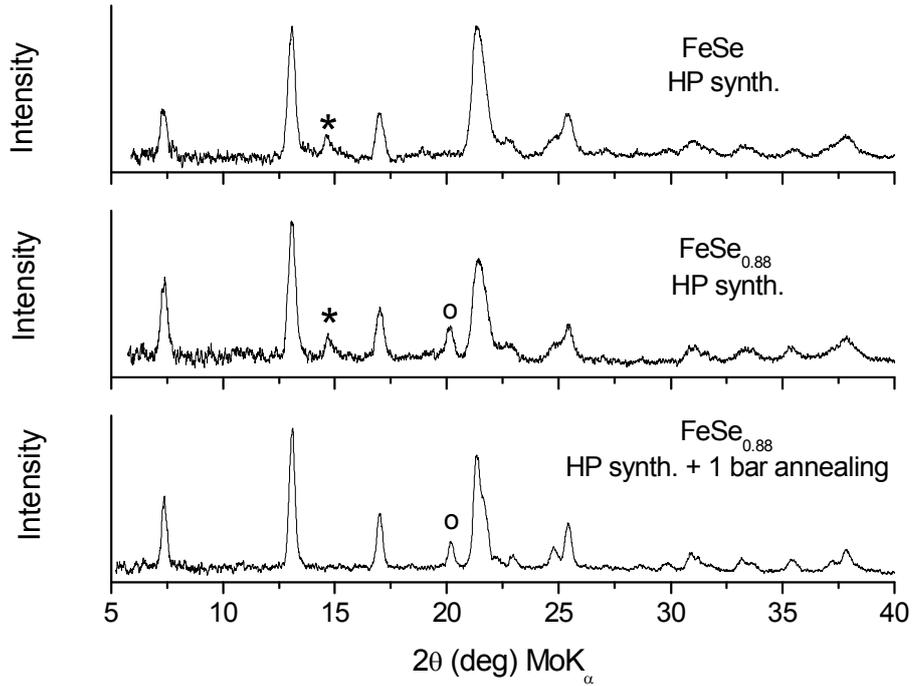

Fig.1 X-ray diffraction patterns of high-pressure synthesized FeSe samples: (*) marks the strongest peak of a hexagonal NiAs-type phase, (o) marks the strongest peak of α-Fe

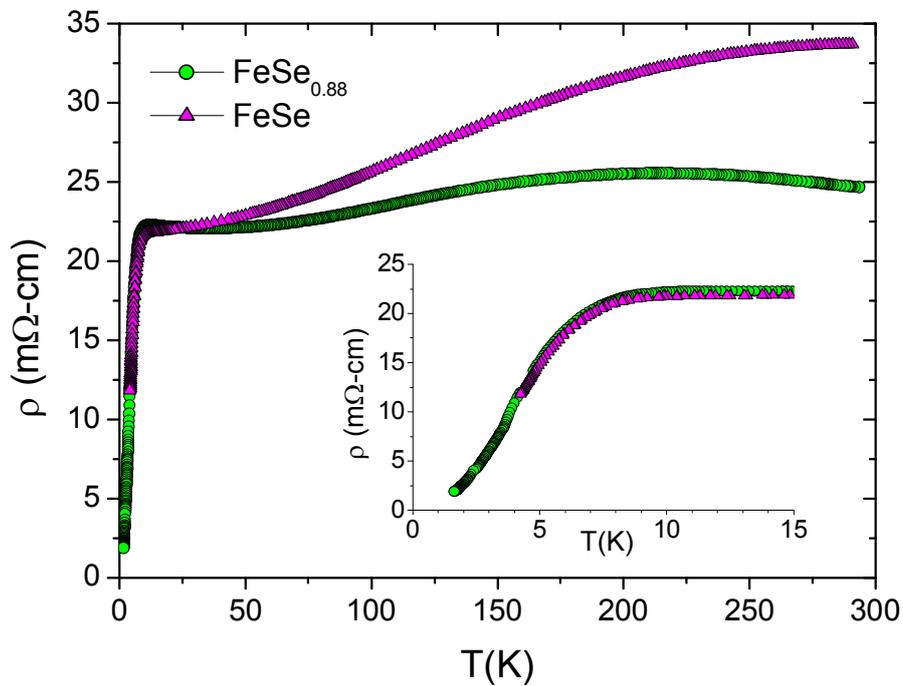

Fig.2 Temperature dependences of the electrical resistivity at normal pressure of two as-prepared FeSe samples (x-ray patterns of these samples are shown in two upper panels of Fig.1)

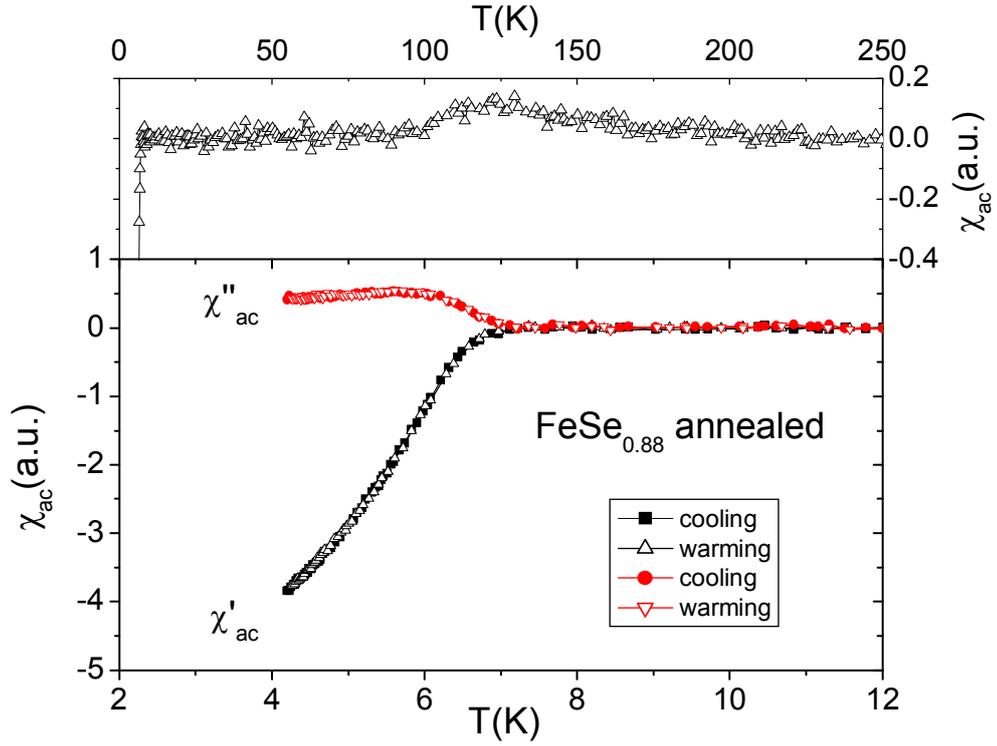

Fig.3 Temperature dependences of ac-susceptibility at normal pressure of the annealed "FeSe$_{0.88}$" sample (x-ray pattern of this sample is shown in the bottom panel of Fig.1)

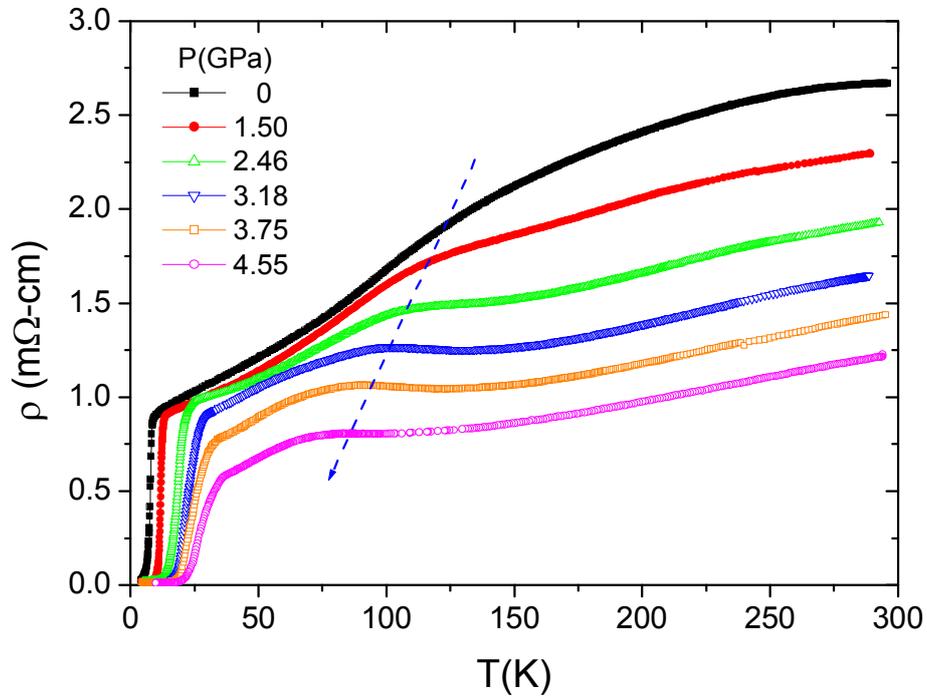

Fig.4 Temperature dependences of the resistivity for the annealed "FeSe$_{0.88}$" sample at different applied pressures. The blue dash arrow indicates the evolution of an SDW-type anomaly in resistivity at high pressure.

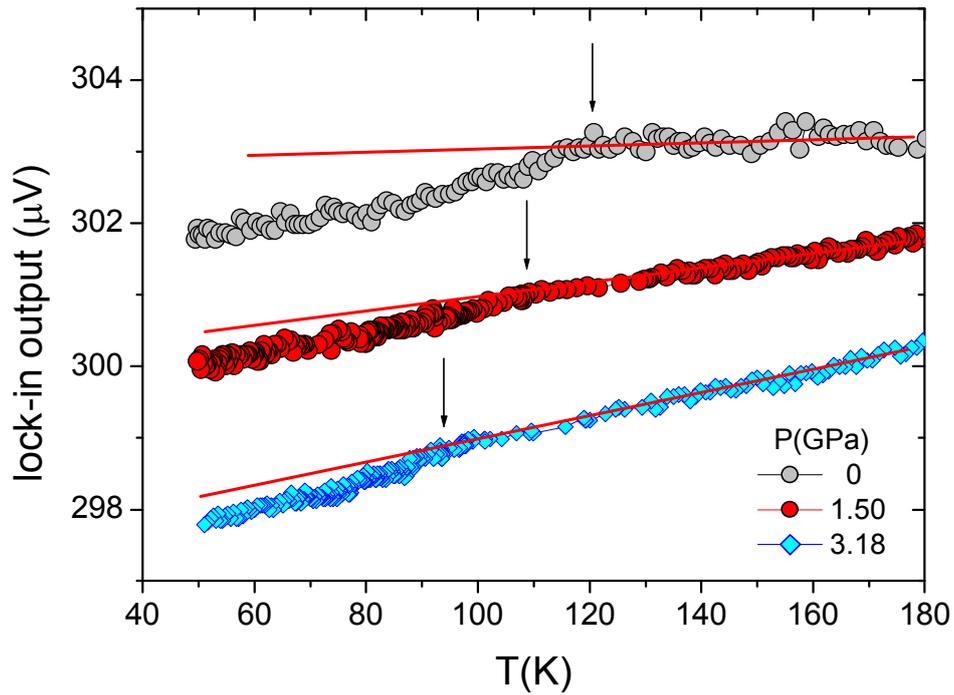

Fig.5 Temperature dependences of the secondary coil output (proportional to susceptibility) at different pressures. The anomalies in ac-susceptibility are marked by arrows.

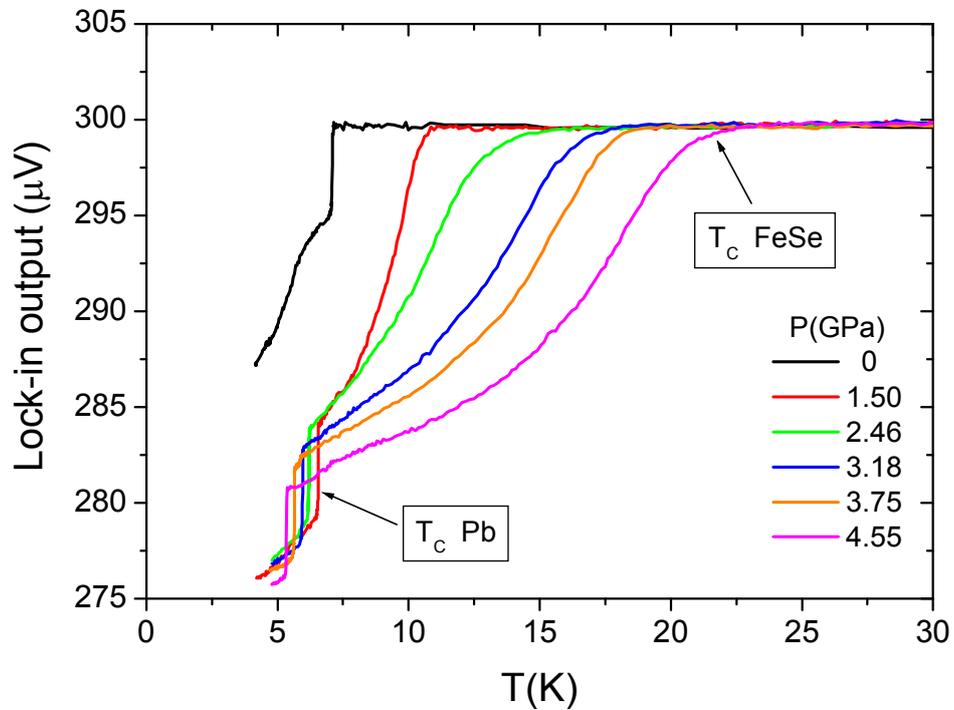

Fig.6 Temperature dependences of the secondary coil output (proportional to susceptibility) at different pressures in the region of superconducting transitions in FeSe and Pb. The superconducting transition in Pb shifts down at high pressure and that of FeSe shifts up at a higher rate.

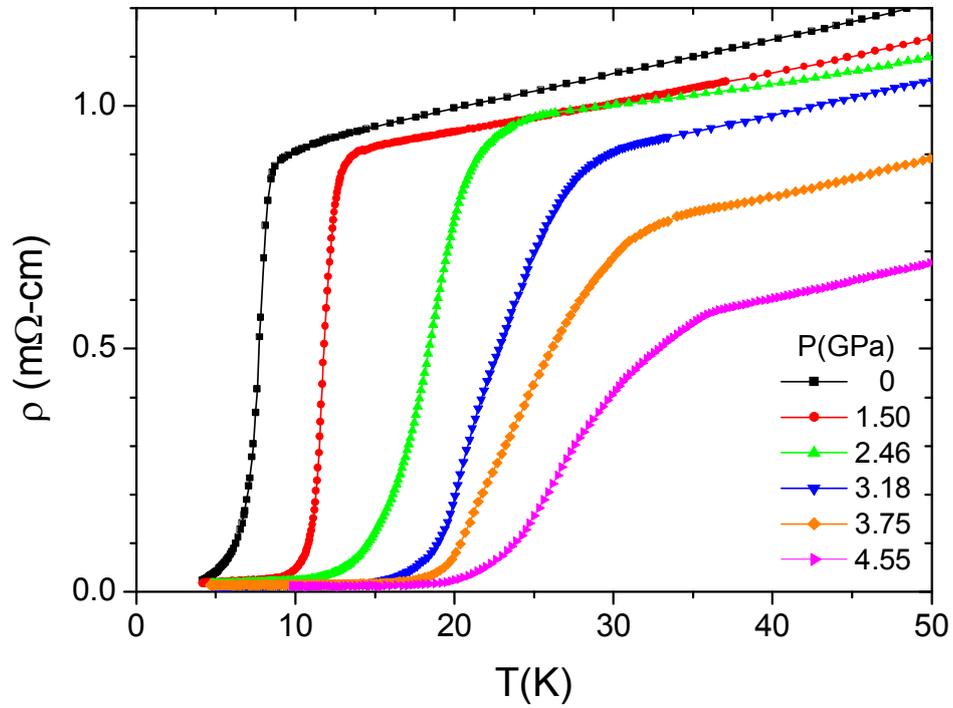

Fig.7 Temperature dependences of the resistivity for the annealed "FeSe$_{0.88}$" sample at different applied pressures in the region of superconducting transitions. Note a giant non-uniform broadening of the superconducting transition above 3 GPa.

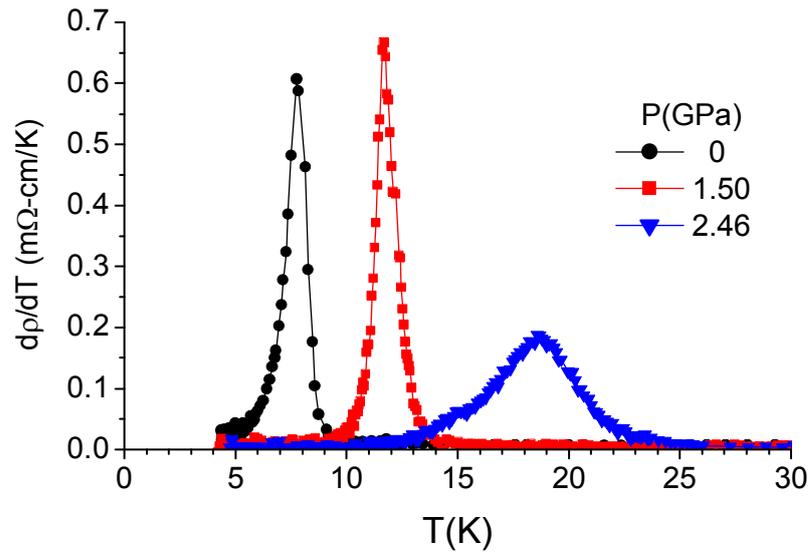
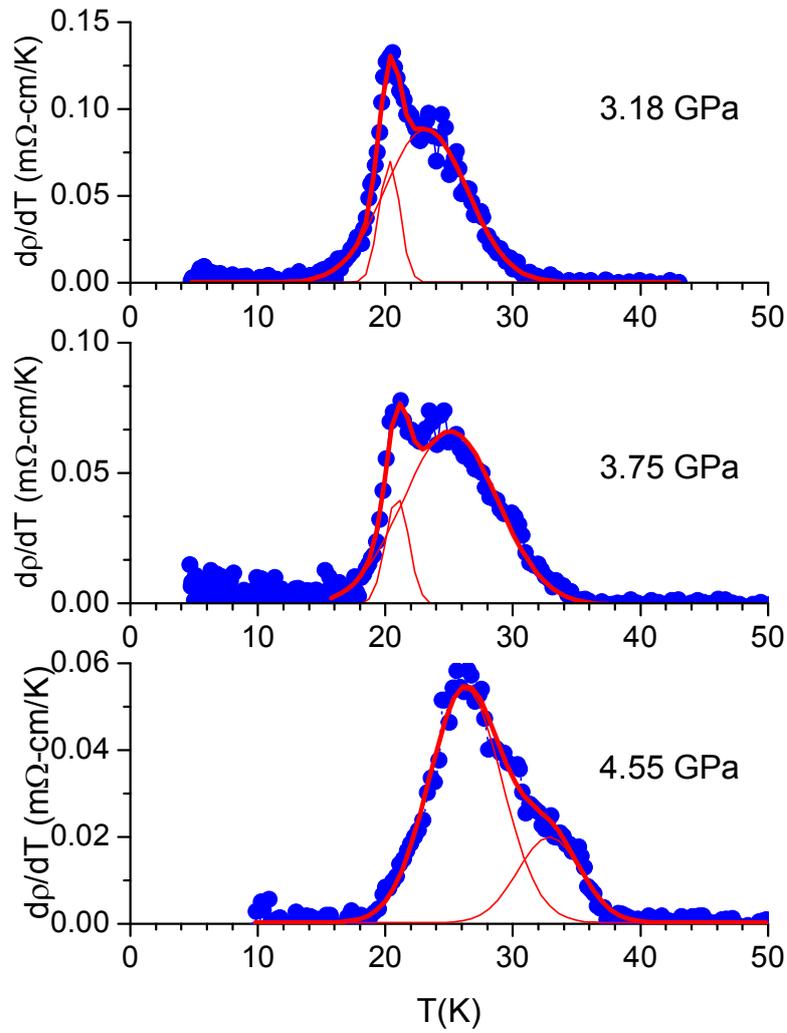

Fig.8 Evolution of the dρ/dT peak at the superconducting transition in FeSe at high pressure. Above 3 GPa two peaks appear; their positions change under pressure at different rates.

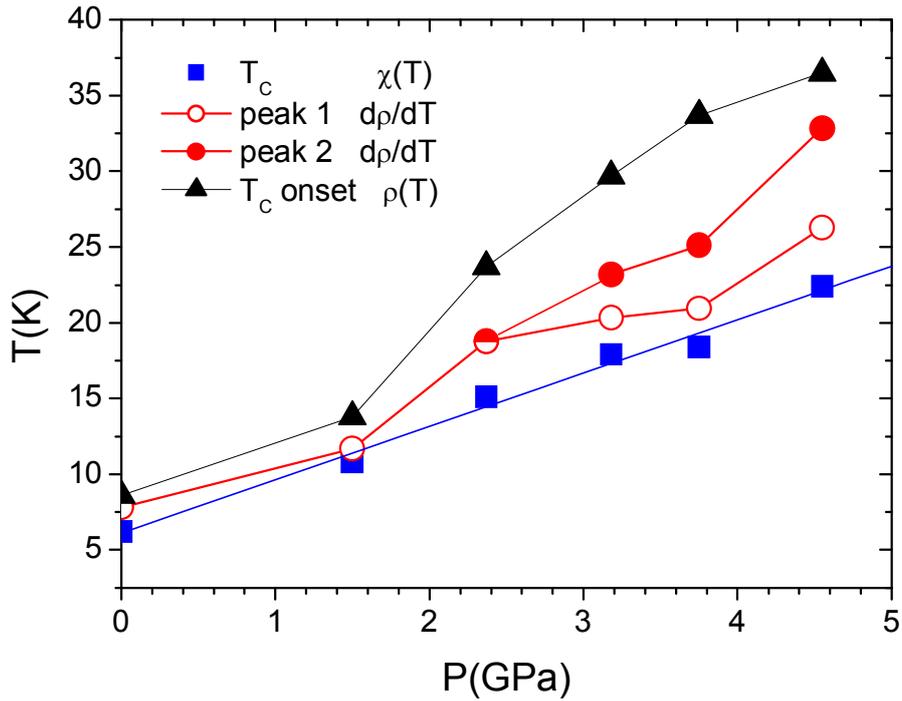

Fig.9 P-T phase diagram of superconductivity in FeSe. Blue squares show the superconducting $T_c$ determined by the appearance of a diamagnetic response in ac-susceptibility; black triangles show the resistive $T_c$ onset temperature. Red circles mark the positions of two peaks in $d\rho/dT$ found in Fig.8.

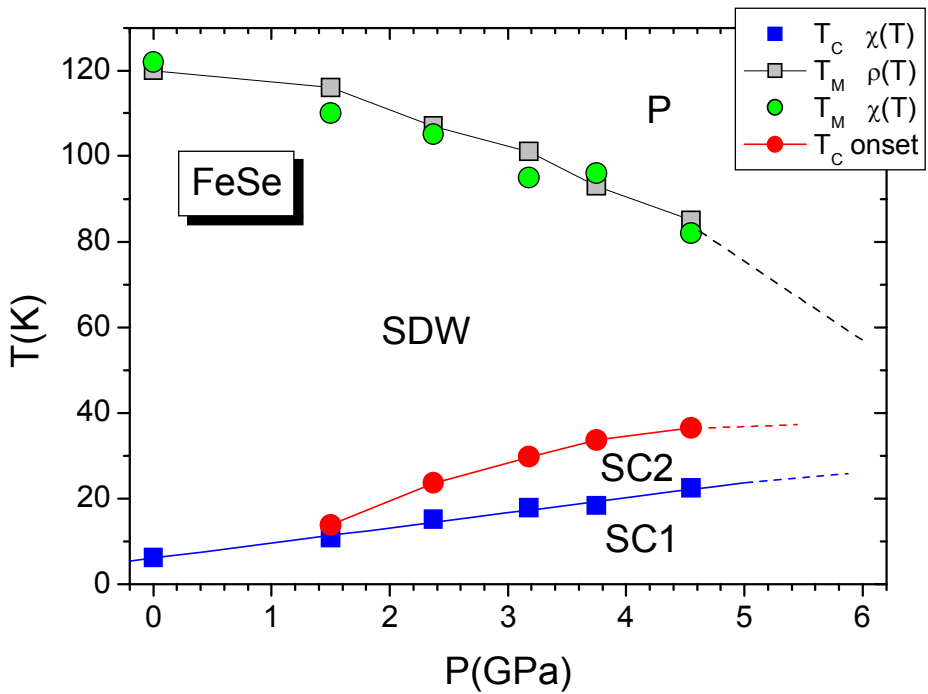

Fig.10 P-T phase diagram of superconductivity and magnetism in FeSe. We explain the appearance of a second superconducting dome by a multiple gap structure of FeSe, originating from multiplicity of its Fermi surface.